\documentclass[lettersize,journal]{IEEEtran}
\usepackage{amsmath,amsfonts}
\usepackage{algorithmicx}
\usepackage{algpseudocode} 

\usepackage{algorithm}
\usepackage{array}
\usepackage[caption=false,font=normalsize,labelfont=sf,textfont=sf]{subfig}
\usepackage{textcomp}
\usepackage{stfloats}
\usepackage{url}
\usepackage{verbatim}
\usepackage{graphicx}
\usepackage{cite}
\usepackage{amsmath}
\usepackage{booktabs} 
\usepackage{threeparttable}
\usepackage{amssymb}
\usepackage{booktabs}
\usepackage{multirow}
\usepackage{makecell}
\usepackage{bbding}  
\usepackage{pifont}
\usepackage{xcolor}
\usepackage{hyperref}
\usepackage{colortbl}

\usepackage{algpseudocode}  
\usepackage{amsmath}  

\definecolor{dark-green}{HTML}{006400}
\definecolor{dark-blue}{HTML}{1976D2}
\definecolor{dark-purple}{HTML}{8d4bbb}
\definecolor{dark-red}{HTML}{D63C3C}
\definecolor{pink}{HTML}{fae6e9}
\definecolor{n1}{HTML}{ff9999}
\definecolor{n2}{HTML}{FFCC99}
\definecolor{n3}{HTML}{FFFF99}
\hypersetup{
    colorlinks=true,
    linkcolor=blue,
    filecolor=magenta,      
    urlcolor=blue,
    citecolor=dark-blue,
}
\newcommand{\shh}[1]{\textcolor{black}{#1}}
\newcommand{\lyb}[1]{\textcolor{black}{#1}}
\begin{document}

\title{M$^2$-ViT: Accelerating Hybrid Vision Transformers with Two-Level Mixed Quantization}

\author{Yanbiao Liang, Huihong Shi, and Zhongfeng Wang,~\IEEEmembership{Fellow,~IEEE}
\thanks{This work was supported by the National Key R\&D Program of China under Grant 2022YFB4400600.}
\thanks{Yanbiao Liang and Huihong Shi are with the School of Electronic Science and Engineering, Nanjing University, Nanjing, China (e-mail: \{{ybliang, shihh}\}@smail.nju.edu.cn).}
\thanks{Zhongfeng Wang is with the School of Electronic Science and Engineering, Nanjing University, and the School of Integrated Circuits, Sun Yat-sen University (email: zfwang@nju.edu.cn).}
\thanks{Correspondence should be addressed to Zhongfeng Wang.}}



\maketitle
 
\begin{abstract}

Although Vision Transformers (ViTs) have achieved significant success, their intensive computations and substantial memory overheads challenge their deployment on edge devices. To address this, efficient ViTs have emerged, typically featuring Convolution-Transformer hybrid architectures to enhance both accuracy and hardware efficiency.
While prior work has explored quantization for efficient ViTs to marry the best of efficient hybrid ViT architectures and quantization, it focuses on uniform quantization and overlooks the potential advantages of mixed quantization.
Meanwhile, although several works have studied mixed quantization for standard ViTs, they are not directly applicable to hybrid ViTs due to their distinct \lyb{algorithmic and hardware characteristics}.
To bridge this gap, we present M$^2$-ViT to accelerate Convolution-Transformer hybrid efficient ViTs with two-level mixed quantization.
Specifically, we introduce a hardware-friendly two-level mixed quantization (M$^2$Q) strategy, characterized by both mixed quantization \textit{precision} and mixed quantization \textit{schemes} (i.e., uniform and power-of-two), to exploit the architectural properties of efficient ViTs. We further build a dedicated accelerator with heterogeneous computing engines to transform our algorithmic benefits into real hardware improvements.
Experimental results validate our effectiveness, showcasing an average of $80\%$ energy-delay product (EDP) saving with comparable quantization accuracy compared to the prior work.

\end{abstract}


\section{Introduction}
\label{sec:intro}
\begin{figure}[!t]
\centering
\setlength{\abovecaptionskip}{0.1cm}
\includegraphics[width=0.7\columnwidth]{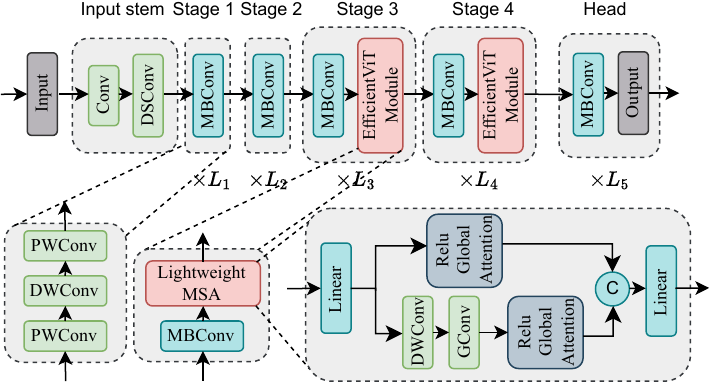}
\caption{{The Convolution-Transformer hybrid architecture of EfficientViT}\cite{cai2022efficientvit}.}
\label{Efficientvit} \vspace{-1.5em}
\end{figure}

\IEEEPARstart{B}{uilt} upon the self-attention mechanism, Vision Transformers (ViTs) have achieved competitive performance in the computer vision \cite{dosovitskiy2021image} and multi-modality \cite{liu2024visual} fields. However, their high computational and memory overheads limit their deployment on resource-constrained edge devices \cite{you2022vitcod}. Particularly, the self-attention mechanism has quadratic computational complexity and is widely recognized as a critical hindrance \cite{you2022vitcod, cai2022efficientvit, Dass2022ViTALiTyUL}.
To solve this limitation, various works have proposed efficient ViTs \cite{cai2022efficientvit,han2023flatten}, which incorporate more efficient attention mechanisms with linear complexity and typically feature Convolution-Transformer hybrid architectures \cite{cai2022efficientvit,han2023flatten}.
For example, as depicted in Fig. \ref{Efficientvit}, the state-of-the-art (SOTA) efficient ViT, dubbed EfficientViT\cite{cai2022efficientvit}, mainly comprises lightweight Multi-Scale Attention (MSA) modules and MBConvs\cite{sandler2019mobilenetv2}, offering much higher accuracy and better hardware efficiency than standard ViTs \cite{dosovitskiy2021image}.


In parallel, {model quantization} approximates floating-point weights/activations using integers, standing out as another effective way to enhance ViTs' efficiency. 
For example, 
FQ-ViT \cite{Lin2021FQViTPQ} identifies quantization challenges within standard ViTs and introduces dedicated approaches to address them.
Furthermore, Trio-ViT\cite{shi2024triovit} delves into the quantization and acceleration for Convolution-Transformer hybrid efficient ViTs, thus advancing the hardware efficiency frontier. Despite Trio-ViT's success in marrying the benefits of both efficient hybrid ViT architectures and quantization, it primarily focuses on uniform quantization and overlooks the potential improvements introduced by mixed quantization. While several works \cite{li2022autovitacc, kloberdanz2023mixquant} have explored mixed quantization, they are dedicated to standard ViTs and not directly applicable to efficient hybrid ViTs, due to their distinct \lyb{algorithmic and hardware characteristics}. 


To close this gap, we propose \textbf{M$^2$-ViT}, targeting the mixed quantization and acceleration for Convolution-Transformer hybrid efficient ViTs, and make the following contributions:


\begin{itemize}
\item{We first introduce a hardware-friendly \textbf{two-level mixed quantization (M$^2$Q)} approach that features both \textbf{\textit{mixed precision}} and \lyb{\textbf{\textit{mixed schemes}} (uniform and power-of-two (PoT))} to fully exploit the architectural properties of efficient ViTs. 
Specifically, (1) for the memory-intensive lightweight layers, 
we investigate the potential of \textbf{\textit{low-bit quantization}} to reduce their bandwidth requirement.
(2) For computation-intensive layers, by analyzing their weight distributions, we explore the potential of \textbf{\textit{PoT quantization}} to boost computational efficiency.
\lyb{Note that we focus on post-training quantization (PTQ) to alleviate the costly fine-tuning process \cite{Lin2021FQViTPQ, shi2024triovit}.}
}
\item{To translate our algorithmic advantages into real hardware efficiency, we develop an \textbf{accelerator} equipped with \textbf{{heterogeneous computing engines}} to accommodate (1) our M$^2$Q strategy with both mixed quantization precision and mixed quantization schemes, and (2) the Convolution- Transformer hybrid architecture inherent in efficient ViTs.}
\item{We conduct experiments to validate our effectiveness. Particularly, compared to the prior work Trio-ViT\cite{shi2024triovit}, we achieve an average $\mathbf{80\%}$ energy-delay product (EDP) saving with comparable ($\downarrow$$0.29\%$) quantization accuracy.}
\end{itemize}

\section{Background}

\subsection{The Structure of EfficientViT}
\label{sec:Background:efficientvit}
As depicted in Fig. \ref{Efficientvit}, EfficientViT has two key attributes: lightweight attention with linear computational complexity to enhance hardware efficiency and Convolution-Transformer hybrid architecture to boost performance. Specifically, EfficientViT primarily comprises MBConvs \cite{sandler2019mobilenetv2} for local information processing and lightweight MSAs for global information extraction.
Each {\bf{MBConv}} consists of two pointwise convolutions (PWConvs) sandwiched by a depthwise convolution (DWConv). 
Besides, the ReLU-based global attention is the core component in each {\bf{lightweight MSA}}. 
This component substitutes the Softmax function in vanilla self-attention with a ReLU-based similarity function, which enables the utilization of the associative property of multiplications to decrease computational complexity from quadratic to linear \cite{cai2022efficientvit}.

\subsection{Uniform Quantization}
Uniform quantization is a basic and most widely adopted quantization method, which converts the floating-point $X$ into $b$-bit integer $X_Q$ as follows:
\begin{equation}\label{uniformquant}
X_Q^{\texttt{Uniform}}=\operatorname{clip}\left(\left\lfloor{\mathrm{X}/}{S}\right\rceil+Z, 0,2^{b}-1\right),
\end{equation}
where $S$ and $Z$ are the scaling factor and zero point, respectively, and they can be determined as follows:
\begin{equation}
\small
S=\frac{\operatorname{max}(X)-\operatorname{min}(X)}{2^{b}-1}, \ Z=\operatorname{clip}\left(\left\lfloor-\frac{\operatorname{min}(X)}{S}\right\rceil, 0,2^{b}-1\right). 
\end{equation}


When performing convolutions, each input channel is processed by the corresponding weight channel within filters and then summed along the channel dimension to produce output. Thus, to eliminate floating-point computations, input activations are generally \textit{layer-wise} quantized, using a common scaling factor for all channels and thus allowing summations to be performed in the integer domain \cite{Lin2021FQViTPQ}.
Similarly, for weights, since summations are constrained to channels within filters, they are typically \textit{filter-wise} quantized, where all weight channels within each filter share the same scaling factor \cite{Lin2021FQViTPQ}.

\section{two-level mixed Quantization (M$^2$Q)}



\begin{figure}[!t]
\centering
\setlength{\abovecaptionskip}{-0.3em}
\includegraphics[width=0.7\columnwidth]{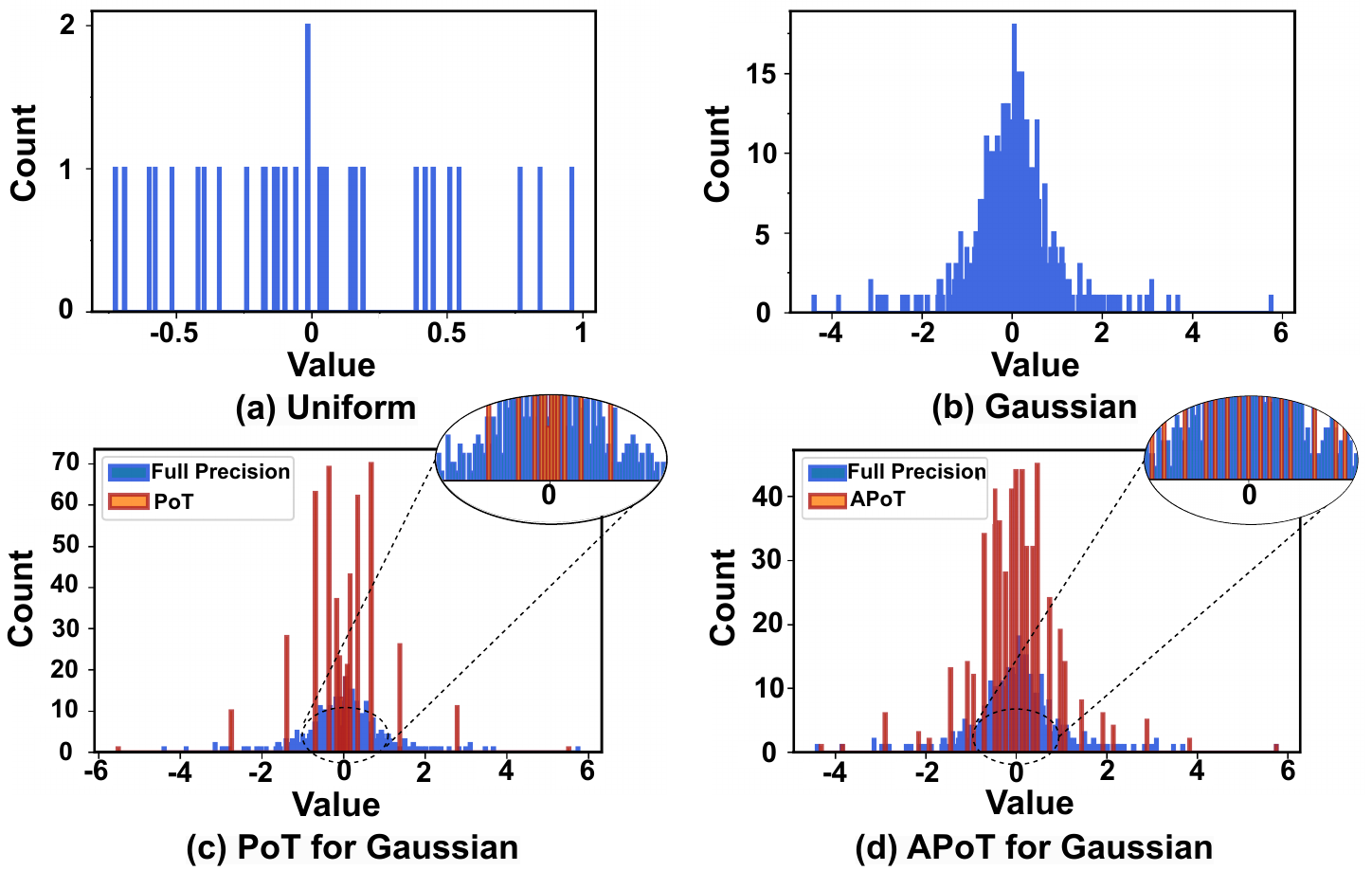}
\caption{Two representative weight distributions for filters in PWConv: (a) Uniform and (b) Gaussian. For filters with Gaussian distribution, different quantization schemes include (c) PoT and (d) APoT.}
\label{fig:distribution} \vspace{-1.5em}
\end{figure}





\subsection{Observations}
\label{sec:quantization_observation}
As shown in Fig. \ref{Efficientvit}, EfficientViT primarily comprises three types of layers: (1) \textit{DWConvs}, where each filter \lyb{(kernel)} has only one channel to process the corresponding input channel for obtaining the output, (2) \textit{PWConvs}, which is equivalent to generic convolutions with $1\times1$ kernel size, and (3) matrix multiplications (\textit{MatMuls}) within lightweight MSAs. \lyb{Based on operation intensity \cite{Williams2008RooflineAI},} we divide these layers into two categories: (1) \textit{computation-intensive layers}, including PWConvs and MatMuls, and (2) \textit{memory-intensive layers}, such as DWConvs, and investigate their algorithmic properties to explore advanced quantization opportunities. 



\emph{1) Computation-Intensive Layers.} To identify potential quantization opportunities, we first choose PWConvs as representative layers and visualize their weight distributions. As seen, the weight distributions across different filters vary from Uniform (Fig. \ref{fig:distribution}a) to Gaussian (Fig. \ref{fig:distribution}b). This observation can be also noted in MatMuls. This variation indicates that merely adopting uniform quantization is sub-optimal and enables the integration of Power-of-Two (PoT) quantization to enhance hardware efficiency. 
Specifically, uniform quantization uniformly distributes quantization bins across all values, making it more appropriate for filters with Uniform distributions.
In contrast, {PoT quantization (see Fig. \ref{fig:distribution}c), which allocates more quantization bins for smaller values, is more suitable for filters with Gaussian distributions. Formally, it is expressed as}:
\begin{equation}\label{eq:pot}
\begin{aligned}
W_{Q}^{{\texttt{PoT}}} = s \times &2^{p},\ \text{where} \\
s = \operatorname{sign}(W), \ p=\operatorname{clip}(\lfloor& log_{2}|{W/}{S}|\rceil,-2^{b}+1,0).
\end{aligned}
\end{equation}
$W$ and $W_{Q}^{\texttt{PoT}}$ are floating-point and PoT quantized weights, respectively. $b$ denotes quantization bit-width. $S$ represents the scaling factor and can be determined by 
$S$=$\operatorname{max}(W)$-$\operatorname{min}(W)$, aiming to re-scale $W$ to [$0$, $1$].
For example, if $W$=$-0.26$, $S$=$2$, $b$=$5$, then $s$=$-1$ and $p$=$-3$.
{{By doing this}, multiplications between activations {$A$} and PoT quantized weights {$W_{Q}^{\texttt{PoT}}$} can be substituted with bitwise shifts, {as formulated in Eq. (\ref{eq:pot_mul})}, thus significantly reducing computational costs.}
\begin{equation}\label{eq:pot_mul}
A \times W_{Q}^{\texttt{PoT}} = A \times s \times 2^{p}=s\times(A>>p).
\end{equation}


\emph{2) Memory-Intensive Layers (DWConvs).} 
Each filter in DWConvs features only one weight channel for handling the corresponding input channel to generate output, which significantly lowers computational costs but limits data reuse opportunities and increases bandwidth demand. Thus, the primary challenge in DWConvs is improving data access efficiency. Fortunately, the small amount of weights per filter of DWConvs inherently offers an opportunity to implement low-bit filter-wise quantization, reducing the bandwidth requirement.

\vspace{-0.8em}
\subsection{\lyb{Two-Level Mixed} Quantization Strategy}
\shh{Motivated by the above observations, we propose a two-level mixed quantization (M$^2$Q) strategy for efficient ViTs\cite{cai2022efficientvit}, which includes (1) mixed-scheme quantization (uniform and PoT) for computation-intensive layers that replace partial multiplications with hardware-efficient bitwise shifts to enhance computational efficiency and (2) mixed-precision quantization to reduce memory access overhead of memory-intensive layers.} Note that our M$^2$Q is exclusively applied to weights, with activations still using standard $8$-bit uniform quantization due to their higher quantization sensitivity \cite{Lin2021FQViTPQ, Li2022IViTIQ}.

\begin{table}[t]
    \centering
    \setlength{\tabcolsep}{4pt}
    \caption{Accuracy of EfficientViT-B1 with weights of computation-intensive layers quantized by different schemes \label{tab:apotpot}} \vspace{-0.8em}
    \renewcommand{\arraystretch}{1.2}
    \resizebox{0.93\linewidth}{!}{
    \begin{threeparttable}{
    \begin{tabular}{c|cccc} \Xhline{3\arrayrulewidth}
         \textbf{EfficientViT-B1\cite{cai2022efficientvit}} & \textbf{Uniform} & \textbf{PoT} & \textbf{APoT}     &  \textbf{APoT\&Uniform} \\ \hline \hline
         \textbf{Top-1 Accuracy* (\%) } &  79.35 & 78.2  & 79.18   &  79.26        \\
         \textbf{Drop (\%)}       &   $\downarrow$0.02    & $\downarrow$1.17 & $\downarrow$0.19  & $\downarrow$0.11 \\ \Xhline{3\arrayrulewidth}
    \end{tabular}}
    \begin{tablenotes}
		\footnotesize
		\item[*] Tested on ImageNet with the input size of $224\times224$ by default.
	  \end{tablenotes} 
    \end{threeparttable}}
    \vspace{-1.7em}
    \label{tab:w8ax}
\end{table}

\begin{table}[t]
    \centering
    \setlength{\tabcolsep}{4pt}
    \caption{Accuracy of EfficientViT-B1 with weights of memory-intensive layers quantized to different bits \label{tab: DW_acc}} \vspace{-0.8em}
    \renewcommand{\arraystretch}{1.2}
    \resizebox{0.93\linewidth}{!}{
    \begin{threeparttable}{
    \begin{tabular}{c|cccccc} \Xhline{3\arrayrulewidth}
         \textbf{EfficientViT-B1\cite{cai2022efficientvit}}  & \textbf{3bit} & \textbf{4bit} & \textbf{5bit} & \textbf{6bit} & \textbf{7bit} & \textbf{8bit}\\ \hline \hline
         \textbf{Top-1 Accuracy* (\%) }  & 79.24  & 79.33 & 79.38 & 79.39 & 79.39 &79.35  \\
         \textbf{Drop (\%)}  & $\downarrow$0.13 & $\downarrow$0.04   & $\uparrow$0.01  & $\uparrow$0.02 & $\uparrow$0.02 & $\downarrow$0.02 \\ \Xhline{3\arrayrulewidth}
    \end{tabular}}
    \begin{tablenotes}
		\footnotesize
		\item[*] Tested on ImageNet with the input size of $224\times224$ by default.
	  \end{tablenotes} 
    \end{threeparttable}}
    \vspace{-2.4em}
    \label{tab:w8ax}
\end{table}

\emph{1) Mixed Quantization Schemes for Computation-Intensive Layers.} As explained in Sec. \ref{sec:quantization_observation}-1), the heterogeneous weight distributions of computation-intensive layers offer us an opportunity to use PoT quantization to reduce computational costs. However, as demonstrated in Table \ref{tab:apotpot}, it inevitably yields accuracy drops compared to $8$-bit uniform quantization. This is because PoT quantization prioritizes accurately representing smaller values near zero, while overlooking bigger values that contribute significantly to final outputs. 
To better balance the major small weights and the minor but important big weights within filters exhibiting Gaussian distributions, {we shift our attention to additive PoT (APoT) quantization\cite{li2020additive}}, which is essentially the combination of two PoT components:
\begin{equation}\label{eq:apot}
W_{Q}^{\texttt{APoT}} = s \times (2^{p_1}+2^{p_2}), \
s = \operatorname{sign}(W),
\end{equation}
{where $p_{1}$/$p_{2}$ are PoT values similarly in Eq. (\ref{eq:pot}).}
By combining two PoT components, APoT strikes a balance between PoT and uniform quantization (see Fig. \ref{fig:distribution}d),
thus enhancing hardware efficiency while maintaining accuracy (see Table \ref{tab:apotpot}).


{After identifying appropriate quantization schemes, the key challenge lies in automatically assigning different quantization schemes to filters with distinct distributions.
\lyb{For example, a PWConv in EfficientViT \cite{cai2022efficientvit} can contain up to $1024$ filters, yielding a design space of $2^{1024}$ for determining schemes for filters in even one layer.} Thus, to enable automatic allocation and save human efforts, we employ the widely adopted Mean Squared Error (MSE) \cite{kloberdanz2023mixquant} to select the optimal quantization scheme that minimizes quantization error for each filter:}
\begin{equation}
\begin{array}{c}
W_{Q} = \underset{W_{Q}}{\arg \min} \: \mathbb{E}\left\|W-W_Q\right\|^2, \\ 
\text { s.t. } W_Q \in\left\{W_Q^{\texttt{APoT}}, W_Q^{\texttt{Uniform}}\right\},
\end{array}
\end{equation}
where $W$ are floating-point and $W_{Q}$ are quantized weights.

\emph{2) Low-Bit Quantization for Memory-Intensive Layers.}
As discussed in Sec. \ref{sec:quantization_observation}-2), the small amount of weights in each filter of DWConvs enable low-bit quantization to reduce bandwidth requirements. 
To determine the optimal bit-width, we quantize DWConvs' weights in EfficientViT from $3$-bit to $8$-bit. As listed in Table \ref{tab: DW_acc}, quantization at $4$-bit or greater yields negligible accuracy drops compared to the {full-precision counterpart}, which supports our low-bit quantization hypothesis. Considering both quantization accuracy and hardware efficiency, we choose $4$-bit filter-wise quantization for DWConvs.

   

\section{M$^2$-ViT's Accelerator}
\label{sec:hw_arch}


The diverse operation types within the hybrid architecture of EfficientViT\cite{cai2022efficientvit} and the mixed quantization precision ($4$bit and $8$bit) and schemes (uniform and APoT) introduced by our M$^2$Q strategy challenge the translation of our algorithmic benefits into real hardware benefits. Specifically, there are mainly three kinds of operations: (a) memory-intensive layers (DWConvs) with $4$-bit uniform quantization; computation-intensive layers (PWConvs/MatMuls) with (b) $8$-bit uniform and (c) APoT quantization. \lyb{However, existing accelerators \cite{shi2024triovit, li2022autovitacc} cannot directly be used due to the lack of tailored computing engines and dataflows, calling for a dedicated accelerator to efficiently perform these operations.}

\begin{figure*}[t]
\centering
\setlength{\abovecaptionskip}{-0.3em}
\includegraphics[width=\linewidth]{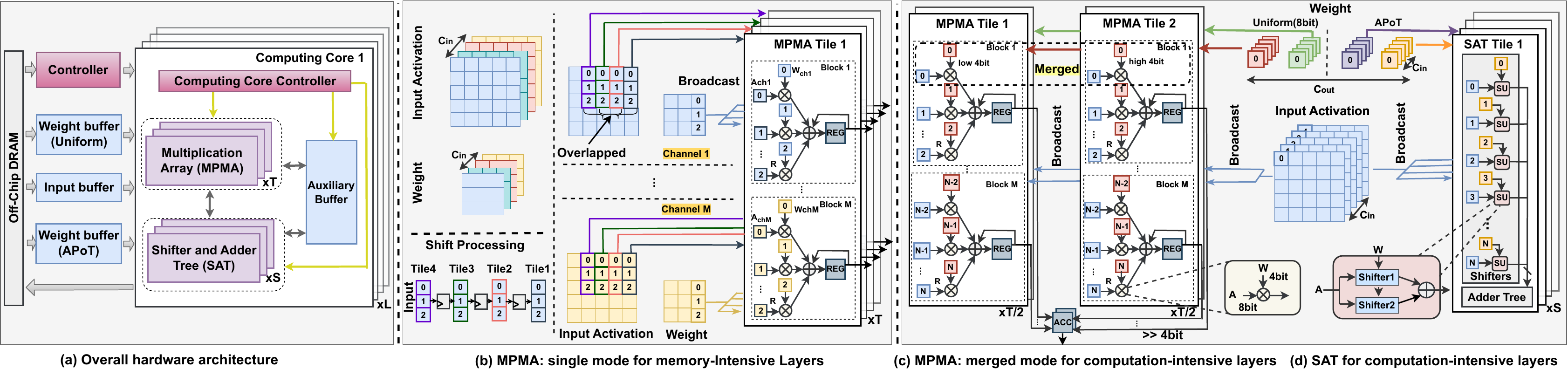}
\caption{(a) Overall hardware architecture, (b) Multiplication Array with single mode and (c) merged mode, and (d) Shifters and Adder Tree.}
\label{fig:hw_arch} \vspace{-1.5em}
\end{figure*}

\textit{\textbf{Overall Hardware Architecture}}. As depicted in Fig. {\ref{fig:hw_arch}a}, our accelerator comprises a controller to provide global control signals, global buffers to store inputs and weights, and $L$ computing cores to process different batches. Specifically, each {computing core} includes a {local controller} to support pre-defined dataflow and an {auxiliary buffer} for caching intermediate results. Besides, it integrates a \emph{Mixed-Precision Multiplication Array (MPMA)}, which features precision-scaleable multipliers to effectively support (a) $4$-bit DWConvs and (b) $8$-bit PWConvs/MatMuls, and a \emph{Shifters and Adder Tree (SAT) engine} equipped with multiple shifters to efficiently handle (c) PWConvs/MatMuls with APoT quantization. We will illustrate the two components - MPMA and SAT, in detail next. 

\emph{1) Mixed-Precision Multiplication Array (MPMA).} 
\lyb{As shown in Fig. \ref{fig:hw_arch}b, the {MPMA} comprises $T$ processing tiles, each containing $M$ PE blocks. Each block features $R$ $4$-bit$\times$$8$-bit (as activations are all represented in $8$-bit) multipliers.} 
To accommodate \lyb{(a) $4$-bit DWConvs and (b) $8$-bit PWConvs/MatMuls}, which differ in both computation patterns and quantization bits, our MPMA is configured to operate in two distinct modes - single mode and merged mode, to respectively handle the aforementioned two types of operations.



\emph{a) Single Mode:} 
As each filter in DWConvs features only one weight channel to handle the corresponding input channel, {input reuse is not available across different filters. To seize the available output reuse and weight reuse opportunities}, we design a {\bf output-parallel} dataflow for DWConvs. 
\lyb{As illustrated in \lyb{Fig. \ref{fig:hw_arch}b}, \lyb{\textbf{\textit{multipliers}}} within each block concurrently process weights from different rows in the same kernel. This allows partial sums to be directly accumulated across cycles within each block, thus enhancing output locality. \lyb{\textbf{\textit{Blocks}}} within the same PE tile independently compute inputs and weights from different channels, enabling parallelism of $M$ in the channel dimension. Meanwhile, \lyb{\textbf{\textit{different PE tiles}}} simultaneously handle inputs from adjacent sliding windows with weights broadcast, facilitating weight reuse and enabling parallelism of $T$ in the output pixel dimension.} 
Additionally, as shown in \lyb{Fig. \ref{fig:hw_arch}b (left)}, by \lyb{simply} using shift registers, we exploit the {reuse of input pixels} from overlaps between adjacent sliding windows. Specifically, during each cycle, different input columns are pipelined and transmitted across PE tiles to generate partial sums for adjacent output pixels, which are then accumulated along cycles to obtain final outputs.


\emph{b) Merged Mode:} 
To support $8$-bit PWConvs and MatMuls (can be treated as PWConvs with large batch size), MPMA operates in merged mode, 
\lyb{where two $4$-bit$\times$$8$-bit multipliers in adjacent PE tiles are merged for $8$-bit$\times$$8$-bit multiplications.}
To facilitate data reuse, we design a {\bf filters-parallel} dataflow. As depicted in \lyb{Fig. \ref{fig:hw_arch}c}, \lyb{\textbf{\textit{all PEs}} within merged PE tiles perform computations along input channel}, providing parallelism of $R \times M$ and enhancing output reuse. Meanwhile, \lyb{\textbf{\textit{different pairs of PE tiles}} parallelly handle different filters with input broadcast}, offering parallelism of $T/2$ along the output channel while enhancing input reuse.


\emph{2) Shifter and Adder Tree (SAT).} 
To support PWConvs/ MatMuls with APoT quantization, we develop a dedicated SAT engine comprising multiple shifters. As shown in  \lyb{Fig. \ref{fig:hw_arch}d}, SAT contains $S$ processing tiles, each containing $N$ shifter units (SU) and an adder tree. \lyb{Each SU comprises two shifters and an adder to support APoT quantization in Eq. (\ref{eq:apot}).} To facilitate data reuse, we propose {\bf filters-parallel} dataflow here, similar to the merged mode of MPMA in Sec. \ref{sec:hw_arch}-1b. \lyb{Specifically, \lyb{\textbf{\textit{all shifter units}}} within the same tile simultaneously process computations along input channel, achieving parallelism of $N$.} The generated partial sums are then directly aggregated by the adder tree to enhance output reuse. \lyb{\lyb{\textbf{\textit{Different tiles}}} handle different filters with inputs broadcast, offering parallelism of $S$ along output channel while facilitating input reuse.}

\textbf{\textit{Execution Flow.}}
As our accelerator integrates heterogeneous computing engines to support different types of operations,
we adopt pipeline processing to enhance hardware utilization. Specifically, (1) for MatMuls with APoT quantization, which are followed by MatMuls with uniform quantization in our algorithm, APoT-quantized MatMuls are first executed by SAT. The generated outputs are immediately sent to MPMA to serve as inputs for subsequent uniform-quantized MatMuls, enabling parallel processing and enhancing hardware utilization.
(2) For DWConvs, which are typically followed by PWConvs in EfficientViT, when MPMA executes DWConvs, the generated outputs are promptly directed to SAT to serve as inputs for the subsequent PWConvs with APoT quantization. Once MPMA finishes DWConvs, it is reallocated to compute filters of PWConvs with uniform quantization, allowing parallel execution with SAT.

\section{Experimental results}
\label{sec:experments}

\subsection{Experimental Setup} 

\textbf{Quantization Setup and Baselines.} 
\textbf{\textit{Quantization Setup:}} Our algorithm is built upon \cite{shi2024triovit} to further explore mixed quantization.
Following \cite{shi2024triovit}, we randomly sample $1024$ images from ImageNet's\cite{Deng2009ImageNetAL} training set as calibration data and test on ImageNet's validation set.
\lyb{To achieve accuracy-efficiency trade-offs, we maintain a $1$:$1$ ratio of APoT to Uniform ($8$-bit) in our mixed-scheme quantization across all computation-intensive layers and align this ratio with the parallelism in corresponding computing engines to enhance hardware utilization.}
\lyb{The quantization is performed offline and the selected schemes are recorded as a series of instructions to guide our accelerator.}
\textit{\textbf{Baselines:}} We compare with Trio-ViT \cite{shi2024triovit}, a SOTA quantization approach dedicated to EfficientViT \cite{cai2022efficientvit} using uniform $8$-bit quantization, \lyb{and Auto-ViT-Acc \cite{li2022autovitacc}, which we apply its mixed quantization method tailored for standard ViTs to EfficientViT \cite{cai2022efficientvit},} in terms of top-$1$ accuracy.

\textbf{Hardware Setup and Baselines.} 
\textbf{\textit{Hardware Setup:}} 
The parallelism of computing cores in our accelerator $(R\times M \times T + N \times S) \times L$ (see Fig. \ref{fig:hw_arch}) is configured to $(3\times 3 \times 16 + 9 \times 8) \times 16$.  To obtain unit energy and overall power, we synthesize our accelerator with Synopsys Design Compiler under a $28$nm TSMC technology and $500$MHz clock frequency. 
\lyb{For fast and accurate estimations, we follow \cite{Dass2022ViTALiTyUL, shi2024triovit} to develop a cycle-level simulator for our accelerator, which utilizes network structure, hardware architecture, dataflow, and technology-dependent unit \lyb{energy} to measure total energy/throughput/latency.}
\textbf{\textit{Baselines:}}
We consider five baselines: (i) full-precision EfficientViT \cite{cai2022efficientvit} (hybrid architecture) executed on the Edge CPU (Qualcomm Snapdragon 8Gen1 CPU); (ii) half-precision EfficientViT executed on the NVIDIA Jetson Nano GPU; (iii) mixed-quantized DeiT \cite{Touvron2020TrainingDI} and (iv) uniform-quantized Swin-T \cite{liu2021swin} respectively executed on their dedicated accelerators \cite{li2022autovitacc, wang2022rowwise}; and (v) uniform-quantized EfficientViT on its dedicated accelerator \cite{shi2024triovit}, which is our most competitive baseline and thus we also implement on ASIC for a fair comparison. We compare them on throughput, energy efficiency, \lyb{end-to-end} latency, \lyb{energy (excluding off-chip memory)}, and energy-delay product (EDP).

\vspace{-1em}
\subsection{Results and Discussions} 
\textbf{Evaluation of M$^2$Q Strategy.} 
As shown in Table \ref{tab:alg_results}, 
thanks to the incorporation of both APoT and low-bit quantization ($4$-bit) in our M$^2$Q strategy, 
we can reduce an average of $\downarrow$$\mathbf{31.5}\%$ computational \lyb{energy} with comparable accuracy (an average of $\downarrow$$\mathbf{0.29}\%$) compared to the SOTA quantization baseline Trio-ViT \cite{shi2024triovit}. \lyb{Besides, M$^2$Q achieves an average \lyb{$\uparrow$$\mathbf{0.95}\%$} accuracy with $10.8\%$ computational energy overhead compared to Auto-ViT-Acc \cite{li2022autovitacc}, showing our superiority. As shown in Table \ref{tab:alg_ablation_study}, MBConvs are more quantization-sensitive than attention but offer higher energy savings. Notably, the overall average accuracy drop is smaller than when only MBConvs are quantized, suggesting the model's robustness and compensation.}


\begin{table}[]
\centering
\caption{TOP-1 accuracy (Acc. \%) and computational \lyb{energy} ($\mu$J) comparisons over Trio-ViT \cite{shi2024triovit} \lyb{and Auto-ViT-Acc\cite{li2022autovitacc}}} \vspace{-1em}
\setlength{\tabcolsep}{0.45em}
\renewcommand{\arraystretch}{1.2}
\resizebox{\linewidth}{!}{
\begin{threeparttable}{
\begin{tabular}{c|c|c|c|c|c|c|c|c|c}
\Xhline{3\arrayrulewidth}
 \hline
 \multicolumn{2}{c|}{\textbf{Model}}         & \multicolumn{2}{c|}{\begin{tabular}[c]{@{}c@{}}\textbf{EfficientViT}\\ \textbf{-B1-R224* \cite{cai2022efficientvit}}\end{tabular}} & \multicolumn{2}{c|}{\begin{tabular}[c]{@{}c@{}}\textbf{EfficientViT}\\ \textbf{-B1-R256 \cite{cai2022efficientvit}}\end{tabular}} & \multicolumn{2}{c|}{\begin{tabular}[c]{@{}c@{}}\textbf{EfficientViT}\\ \textbf{-B1-R288 \cite{cai2022efficientvit}}\end{tabular}} & \multicolumn{2}{c}{\begin{tabular}[c]{@{}c@{}}\textbf{EfficientViT}\\ \textbf{-B2-R224 \cite{cai2022efficientvit}}\end{tabular}} \\ \hline \hline 
 \textbf{Method}& \textbf{W/A} & \textbf{Acc.}& \textbf{\lyb{Energy}}& \textbf{Acc.}&\textbf{\lyb{Energy}}&\textbf{Acc.}&\textbf{\lyb{Energy}}&\textbf{Acc.}&\textbf{\lyb{Energy}}
 
\\ \hline 
 \textbf{Full Precision} &32/32& 79.39  & 747.94                                                         & 79.92 & 976.9                                                          & 80.41 & 1236.38                                                          & 82.10 & 2313.13                                                                                                                    \\ \hline
 \textbf{Trio-ViT\cite{shi2024triovit}} &8/8     & 78.64 & 26.06                                                      & 78.93 & 34.03                                                          & 79.58   & 43.07                                                    & 80.97 &80.58 \\ \hline
  \lyb{\textbf{Auto-ViT-Acc\cite{li2022autovitacc}}} & \lyb{M/8}     & \lyb{76.92} & \lyb{16.13}  & \lyb{77.80} & \lyb{21.07}    & \lyb{78.00}   &  \lyb{26.66}  & \lyb{80.42} & \lyb{49.88} \\ \hline
\rowcolor{dark-green!16} \textbf{Ours} &M/8 & \textbf{78.45} & \textbf{17.85}        & \textbf{78.78} & \textbf{23.31}                                                            & \textbf{79.15} & \textbf{29.5}                                                            & \textbf{80.56} & \textbf{55.64}   

\\ \Xhline{3\arrayrulewidth}
\end{tabular}} 
\begin{tablenotes}
		\footnotesize
		\item[*] R224 denotes the resolution of input images is $224\times224$, and so on.
	  \end{tablenotes} 
\end{threeparttable}}\label{tab:alg_results} \vspace{-1.5em}
\end{table}

\begin{table}[!t]
    \centering
    \setlength{\tabcolsep}{4pt}
    \caption{\lyb{Ablation study of M$^2$Q strategy on different layers}} \vspace{-1em}
    \renewcommand{\arraystretch}{1.2}
    \resizebox{\linewidth}{!}{
    \begin{threeparttable}{
    \begin{tabular}{c|c|c|c|c|c|c|c|c|c} \Xhline{3\arrayrulewidth}
         \multicolumn{2}{c|}{\textbf{Model}}  & \multicolumn{2}{c|}{\begin{tabular}[c]{@{}c@{}}\textbf{EfficientViT}\\ \textbf{-B1-R224* \cite{cai2022efficientvit}}\end{tabular}} & \multicolumn{2}{c|}{\begin{tabular}[c]{@{}c@{}}\textbf{EfficientViT}\\ \textbf{-B1-R256 \cite{cai2022efficientvit}}\end{tabular}} & \multicolumn{2}{c|}{\begin{tabular}[c]{@{}c@{}}\textbf{EfficientViT}\\ \textbf{-B1-R288 \cite{cai2022efficientvit}}\end{tabular}} & \multicolumn{2}{c}{\begin{tabular}[c]{@{}c@{}}\textbf{EfficientViT}\\ \textbf{-B2-R224 \cite{cai2022efficientvit}}\end{tabular}} \\ \hline \hline

         \textbf{Layers} & W/A & \textbf{Acc.}& \textbf{Energy}& \textbf{Acc.}&\textbf{Energy}&\textbf{Acc.}&\textbf{Energy}&\textbf{Acc.}&\textbf{Energy} \\ \hline     
            
         \textbf{All} & 8/8 & 78.64 & 26.06   & 78.93  & 34.03  & 79.58 & 43.07 & 80.97 & 80.58 \\ \hline  
         
         \textbf{FFN(MBConv)}  & M/8  & 78.33 & 20.34 & 78.48 & 26.56 & 79.44 & 33.61 & 80.37 & 63.19  \\ \hline
         
         \textbf{Attention}  & M/8 & 78.55 & 23.76  & 79.00  & 31.04 & 79.56 & 39.29 & 80.55 & 73.35 \\ \hline 
         
         \textbf{All} & M/8 & 78.45  & 17.85  & 78.78  & 23.31  & 79.15  & 29.50 & 80.56  & 55.64   
         
         \\ \Xhline{3\arrayrulewidth}
    \end{tabular}}
    \begin{tablenotes}
		\footnotesize
		\item[*] R224 denotes the resolution of input images is 224 × 224, and so on.
	  \end{tablenotes} 
    \end{threeparttable}}
    \vspace{-1.5em}
    \label{tab:alg_ablation_study}
\end{table}

\begin{table}[!t]
\centering
\setlength{\tabcolsep}{1.pt}
\caption{Comparisons with SOTA transformer accelerators} \vspace{-1em}
\renewcommand{\arraystretch}{1.2}
\resizebox{\linewidth}{!}{
\begin{tabular}{c|c|c|c|c|c|c|c}
\Xhline{3\arrayrulewidth}
\textbf{Accelerator} &                                                            \multicolumn{2}{c|}{\textbf{\begin{tabular}[c]{@{}c@{}}General Computing \\ Platforms (CPU/GPU) \end{tabular}}}&   \textbf{\begin{tabular}[c]{@{}c@{}}Auto-ViT \\ -Acc \cite{li2022autovitacc} \end{tabular}}  &   \textbf{\begin{tabular}[c]{@{}c@{}}ViT \\ Accelerator \cite{wang2022rowwise} \end{tabular}}  & \multicolumn{2}{c|}{\textbf{Trio-ViT \cite{shi2024triovit}} }   & \textbf{\textcolor{dark-green} {Ours}}  \\ \hline \hline
\textbf{\begin{tabular}[c]{@{}c@{}}Device/ \\ Technology \end{tabular}}  & \textbf{CPU} & \textbf{\begin{tabular}[c]{@{}c@{}}NVIDIA \\ Jetson Nano \end{tabular}}& \textbf{\begin{tabular}[c]{@{}c@{}}Xilinx \\ ZCU102 \end{tabular}}& \textbf{\begin{tabular}[c]{@{}c@{}}TSMC \\ 40nm \end{tabular}} & \textbf{\begin{tabular}[c]{@{}c@{}}Xilinx \\ ZCU102 \end{tabular}} & \textbf{\begin{tabular}[c]{@{}c@{}}TSMC \\ 28nm \end{tabular}}  & \textbf{\begin{tabular}[c]{@{}c@{}}TSMC \\ 28nm \end{tabular}}  \\ \hline
\textbf{\begin{tabular}[c]{@{}c@{}}Frequency\\ (GHz)\end{tabular}} & \textbf{1.8-3.0} & \textbf{0.9}& \textbf{0.15} & \textbf{0.6}  & \textbf{0.2} & \textbf{0.5} & \textbf{0.5} \\ \hline
\textbf{Format} & \textbf{FP32} & \textbf{FP16}    & \textbf{Mixed}& \textbf{INT8}& \textbf{INT8} & \textbf{INT8}    & \textbf{Mixed}      

\\ \hline
\textbf{Model} & \begin{tabular}[c]{@{}c@{}}\textbf{EfficientViT}\\ \textbf{-B1-R224 \cite{cai2022efficientvit}}\end{tabular} & \begin{tabular}[c]{@{}c@{}}\textbf{EfficientViT}\\ \textbf{-B1-R224 }\end{tabular}     & \begin{tabular}[c]{@{}c@{}}\textbf{DeiT-}\\ \textbf{small \cite{Touvron2020TrainingDI}}\end{tabular}& \textbf{Swin-T\cite{liu2021swin}}& \begin{tabular}[c]{@{}c@{}}\textbf{EfficientViT}\\ \textbf{-B1-R224 }\end{tabular}    & \begin{tabular}[c]{@{}c@{}}\textbf{EfficientViT}\\ \textbf{-B1-R224 }\end{tabular}& \begin{tabular}[c]{@{}c@{}}\textbf{EfficientViT}\\ \textbf{-B1-R224 }\end{tabular}

\\ 
\hline
\textbf{GFLOPs} & {0.52}   & {0.52} & {4.6}  &  {4.5} &  {0.52}   & {0.52}  & \textbf{0.52}      

\\ 
\hline
\textbf{\begin{tabular}[c]{@{}c@{}}Throughput\\ (GOPS)\end{tabular}} & 54.7  &41.9 & 1418  &  430 (peak) & 791  &  1978     & \textbf{2150}      \\ 
\hline
\textbf{\begin{tabular}[c]{@{}c@{}}Energy Efficiency\\ (GOPS/W)\end{tabular}} & 5.0 & 4.2   & 137.2 & \textbf{-} & 108 & 757.9   & \textbf{2687.5}      \\\hline 
\textbf{Latency (ms)} & 19 & 24.8& 6.4  & 22.4 & 1.31 & 0.53    & \textbf{0.48}  
\\ \hline 
\lyb{\textbf{Energy (mJ)}}   & \lyb{\textbf{-}} & \lyb{\textbf{-}} & \lyb{\textbf{-}}    & \lyb{\textbf{-}}    & \lyb{\textbf{-}} & \lyb{8.11} & \lyb{\textbf{1.83}}   
\\ \hline 
\textbf{EDP (mJ$\cdot$ms)}   & \textbf{-} & \textbf{-} & \textbf{-}    & \textbf{-}    & \textbf{-} & 4.3 & \textbf{0.88}   \\ 

 \Xhline{3\arrayrulewidth}                                                             
\end{tabular}} \label{tab:hw_results} \vspace{-1.7em}
\end{table}

\textbf{Evaluation of M$^2$-ViT's Accelerator.} 
From Table \ref{tab:hw_results} we can see that: 
 \textbf{(i)} Compared to \lyb{EfficientViT-B1-R224}\cite{cai2022efficientvit} executed on CPU/GPU, we achieve $\uparrow$$\mathbf{39.3}\times$$\sim$$\uparrow$$\mathbf{51.3}\times$ throughput and $\uparrow$$\mathbf{537.5}\times$$\sim$$\uparrow$$\mathbf{639.9}\times$ energy efficiency. \textbf{(ii)} When compared with mixed-quantized DeiT \cite{Touvron2020TrainingDI}  and uniform-quantized Swin-T \cite{liu2021swin} on their dedicated accelerators \cite{li2022autovitacc,wang2022rowwise}, we offer  $\uparrow$$\mathbf{1.52}\times$ and $\uparrow$$\mathbf{5.00}\times$ throughput, respectively. \textbf{(iii)} As for our most competitive baseline, uniform-quantized EfficientViT on its accelerator \cite{shi2024triovit} and implemented on the same ASIC technology as us, we can gain $\uparrow$$\mathbf{3.55}\times$ energy efficiency, \lyb{$\mathbf{77}\%$ energy reduction} and $\mathbf{80}\%$ EDP saving.
 \shh{As shown in Table \ref{tab:area_and_power_comparison}, our throughput/latency benefits mainly from the area efficiency of our computing units, enabling higher parallelism. The energy/power savings stem from our hardware-efficient computing units and low buffer overhead, which are attributed to the low bit-widths introduced by our quantization scheme.}

\begin{table}[!t]
\centering
\caption{\centering{\lyb{Comparisons between computing units and buffers}}} \vspace{-1em}
\setlength{\tabcolsep}{1.pt}
\renewcommand{\arraystretch}{1.2}
\resizebox{\linewidth}{!}{
\begin{threeparttable}{
\begin{tabular}{c|c|c|c|c|c}
\Xhline{3\arrayrulewidth}
 \hline
  \multicolumn{3}{c|}{\textbf{8-bit Uniform Quantization Trio-ViT \cite{shi2024triovit}}} & \multicolumn{3}{c}{\textbf{\textcolor{dark-green} {Ours}}} 
 
 \\ \hline \hline 
 \textbf{Computing Unit}& \textbf{Area(mm$^2$)} & \textbf{Power(mW)}& \textbf{Computing Unit}& \textbf{Area(mm$^2$)} & \textbf{Power(mW)}
 
\\ \hline 
 \textbf{8bit$\times$8bit Mul.} &141.87& 2.63$\times$10$^{-2}$  &\textbf{\begin{tabular}[c]{@{}c@{}}Precision-Scaleable Mul. \\ (8$\times$8bit or two 4$\times$8bit) \end{tabular}
 }                                                        & 145
 & 2.54$\times$10$^{-2}$                                                                                                                                                                         \\ \hline
 \textbf{-} &   -  &  - &\textbf{Shifter Unit}                                                     & 71 & 1.06$\times$10$^{-2}$                                                           

\\ \hline \hline
\textbf{Weight Buffer}& \textbf{Area(mm$^2$)} & \textbf{Power(mW)}& \textbf{Weight Buffer}& \textbf{Area(mm$^2$)} & \textbf{Power(mW)}   
\\ \hline                               \textbf{Buffer (8bit)} &   128144  &  18.0256 &\textbf{Buffer (4bit)}                                                     & 83472 & 11.8784                                 
 \\ \hline                               \textbf{Buffer (8bit)} &   61692  &  8.5676 &\textbf{Buffer (APoT)}                                                     & 45448 & 6.3228

  \\ \hline                               \textbf{Total} &   189836  &  26.5932 &\textbf{Total}                                                     & 128920 & 18.2012

\\ \Xhline{3\arrayrulewidth}
\end{tabular}}

\end{threeparttable}}\label{tab:area_and_power_comparison} \vspace{-2.0em}
\end{table}

\label{sec:hw_results}

\section{Conclusion}


In this brief, we propose M$^2$-ViT to marry the hardware efficiency of both quantization and hybrid Vision Transformers (ViTs). We first develop a two-level mixed quantization (M$^2$Q) strategy upon the SOTA EfficientViT \cite{cai2022efficientvit} to fully exploit its architectural properties. To translate our algorithmic advantages into real hardware efficiency, we further design a dedicated accelerator with heterogeneous computing engines. We finally conduct experiments to validate our effectiveness.



\bibliographystyle{ieeetr}

\bibliography{main}

\end{document}